\documentclass[9pt]{article}

\usepackage{xcolor}
\usepackage{color, colortbl}
\usepackage{PRIMEarxiv}

\usepackage[utf8]{inputenc} 
\usepackage[T1]{fontenc}    
\usepackage{hyperref}       
\usepackage{url}            
\usepackage{booktabs}       
\usepackage{amsfonts}       
\usepackage{nicefrac}       
\usepackage{microtype}      
\usepackage{lipsum}
\usepackage{fancyhdr}       
\usepackage{graphicx}       
\graphicspath{{media/}}     

\usepackage{titlesec}
\titleformat{\subsubsection}
  {\normalfont\normalsize\bfseries\itshape}{\thesubsubsection.}{1em}{}

\title{Developing Compelling Safety Cases
}

\author{
  Richard Hawkins \\
  University of York \\
  \texttt{richard.hawkins@york.ac.uk} \\
}

\begin{document}
\maketitle

\begin{abstract}

This paper describes a method for creating compelling safety cases. The method seeks to help improve safety case practice in order to address the weaknesses identified in current practice, in particular confirmation bias, after-the-fact assurance and safety cases as a paperwork exercise. Rather than creating new notations and tools to address these issues, we contend that it is improvements in the safety case process that will make the most significant improvement to safety case practice. Our method builds upon established approaches and best practice to create an approach that will ensure safety cases are risk-focussed, seek to identify ways in which the system may \textbf{\textit{not}} be safe (rather than just assuming it is), drive safe design and operation of the system (influencing the system itself rather than just documenting what's there), are used to support decisions made throughout the life of the system, including system operation and change, and encourage developers and operators to think about and understand why their system is safe (and when it isn't). A simple example of an infusion pump system is used to illustrate how the new method is applied in practice. 

\end{abstract}

\section{Introduction}
\label{sec:intro}

Safety cases provide a way to communicate a clear, comprehensive and defensible argument that a system is acceptably safe to operate \cite{kelly1999arguing}. They consist of a structured argument, supported by a rigorous body of evidence \cite{0056}. The concept and use of safety cases developed in response to a number of high profile disasters, notably the Windscale reactor fire in 1957 \cite{penney2017report}, the Flixborough factory explosion in 1974 \cite{flix} and the Piper Alpha oil platform fire of 1988 \cite{cullen1993public}, where the requirement for a safety case to explicitly argue the safety of the system was recommended as a way to help prevent similar catastrophes from occurring. The adoption of safefty cases has since become widespread across many different industries including rail \cite{std2019railway}, automotive \cite{26262}, nuclear \cite{onr}, air \cite{icao}, medical \cite{tir382019medical}, oil and gas \cite{hse2006offshore} and defence \cite{mod}.

\begin{figure*}
\centering
\includegraphics[width=0.4\textwidth]{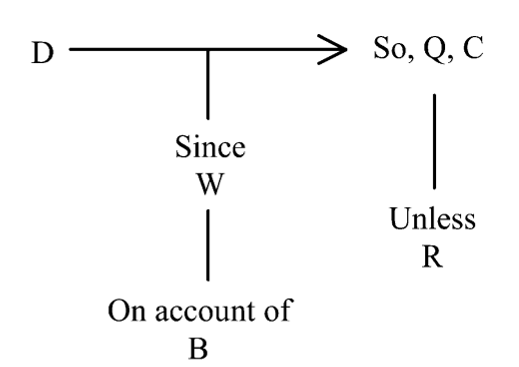}
\caption{Toulmin's model of inductive arguments (from \cite{toulmin2003uses}).}
\label{fig:toulmin}
\end{figure*}

In order to support the development of compelling safety cases, a number of safety argument notations have been developed to help to create and document clear safety argument structures. The most commonly used safety argument notation is the Goal Structuring Notation (GSN) \cite{group2021a}, with others such as Claims, Argument, Evidence (CAE) \cite{cae} also being used. These notations are also supported by methods that provide guidance on how to use the notation to create safety arguments (such as the 6-step method for GSN \cite{kelly1999arguing}). All safety argument notations build upon the same foundations of inductive argumentation as described by Stephen Toulmin in the 1950s \cite{toulmin2003uses}. Safety arguments presented in safety cases for systems are mostly inductive in nature, where the truth of the safety claims cannot be proven with absolute certainty, and are instead supported by the argument and evidence with a level of probability. Crucially, Toulmin's approach involves assessing the argument being made in order to make it stronger (a theme we return to later). Toulmin introduced the core elements required of any inductive argument, and these elements remain crucial to the arguments used in safety case to this day. Figure \ref{fig:toulmin} shows these elements and their relationships:
\begin{itemize}
    \item Data (D): The evidence, information, etc. which support the truth of the claim. 
    \item Claim (C): The conclusion of the argument. This may be an intermediate step used as data for the next inference.
    \item Warrant (W): The reasons that link the data to the claim. 
    \item Backing (B): Statements that serve as evidence for showing that the warrants are true.
    \item Rebuttals (R): Exceptions to the claim, showing circumstances when the claim might not be true.
    \item Qualifiers (Q): Statements or phrases which reflect the level of probability or truth of the claim.
\end{itemize}

Despite the increasing adoption of safety cases across different industries and the establishment of safety argument notations and techniques, it appears that the standard of safety case practice remains mixed. Whilst there are clearly examples of good practice across industry, it is also clear that there are many more examples where safety cases are not being used effectively. This was brought into sharpest focus in the report by Charles Haddon-Cave into the 2006 Nimrod accident in Afghanistan \cite{qc2009nimrod}. Amongst many findings, the report considered the role of the safety case that had been developed for the Nimrod aircraft. Haddon-Cave determined that \textit{``if the Nimrod Safety Case had been prepared with proper skill, care and attention, ... XV230 would not have been lost.''} This supports the assertion that safety cases are crucial to ensuring the safety of systems, but only if they are used effectively. Haddon-Cave goes on to say that, \textit{``Unfortunately, the Nimrod Safety Case was a lamentable job from start to finish. It was riddled with errors. It missed the key dangers. It was essentially a ‘paperwork’ exercise. It was virtually worthless as a safety tool''}.  In particular, he highlights the need to \textit{``...focus attention on the fact that [safety cases] are about managing risk, not assuming safety''}.

There are other sources of criticism of how safety cases are being used in practice \cite{kelly2008safety}, \cite{habli2021safety}. The most high profile of these is probably Nancy Leveson, who has published a number of papers that critique the safety case approach \cite{leveson2020white}, \cite{leveson2011use}. In particular Leveson identifies a number of problems she perceives with the use of safety cases that echo the key problems highlighted by Haddon-Cave. These are:
\begin{itemize}
    \item \textbf{Confirmation bias} - seeking out evidence to support a pre-determined conclusion whilst tending to ignore information that may be contradictory. It is often stated that current safety case practice encourages confirmation bias by starting from the presumption that a system is sufficiently safe. A good safety case requires that focus is also paid to understanding why a system may be unsafe.
    \item \textbf{After-the-fact assurance} - attempting to provide assurance of the safety of a system once development of the system is complete. Current practice can sometimes view the creation of a safety case as an activity to be performed at the end of the lifecycle, thus offering no opportunity for the development of the safety case to influence the design of the system in a timely manner. This practice can also increase the likelihood of confirmation bias in a safety case since it increases pressure to ``show the system is safe''.
    \item \textbf{Paperwork exercise} - viewing the task of creating a safety case as purely one of creating a document. If safety cases are viewed simply as a paperwork exercise (to show compliance with a requirement to produce a document) then they are of little value in terms of ensuring a system is safe to operate. Safety cases treated in this way will not play an active role in the development and assurance of safe systems, and are instead likely to become ``shelf-ware''. They are also likely to produced after-the-fact and subject to confirmation bias. 
\end{itemize}

There have been many papers written discussing how these safety case challenges could be addressed. These often focus on new notations, such as \cite{goodenough2015eliminative}, \cite{holloway2023friendly}, or increased tool support and automation, such as \cite{denney2012advocate}, \cite{matsuno2010dependability}. In contrast, the premise of this paper is that to make a substantive improvement in safety case practice, additional notations and tools are not required, existing notations, building on Toulmin's foundations, provide everything that is needed to create compelling safety cases (as is seen in good practice today). Instead we believe that the solution lies in better application of the existing concepts to address the identified challenges and to better support safety case development for ever more complex systems and emerging classes of system with more rapid development and update cycles. Again we can find support for this in the Nimrod report where it highlights that, \textit{``The basic aims, purpose and underlying philosophy of Safety Cases were clearly defined, but there was limited practical guidance as to how, in fact, to go about constructing a Safety Case''} \cite{qc2009nimrod}.

In this paper we propose a safety case development method that builds on establishes approaches in order to help improve safety case practice. The aim of the method is to address the challenges identified above in order to create safety cases that:
\begin{itemize}
    \item are risk-focussed
    \item seek to identify ways in which the system may \textbf{\textit{not}} be safe (rather than just assuming it is)
    \item drive safe design and operation of the system (influencing the system itself rather than just documenting what's there) \cite{graydon2007assurance}
    \item are used to support decisions made throughout the life of the system, including system operation and change \cite{kelly2001systematic}
    \item encourage developers and operators to think about and understand why their system is safe (and when it isn't)
\end{itemize}

The paper describes the method and illustrates its application through the use of an example.

\section{Proposed Method}
\label{sec:method}

\begin{figure*}
\includegraphics[width=1\textwidth]{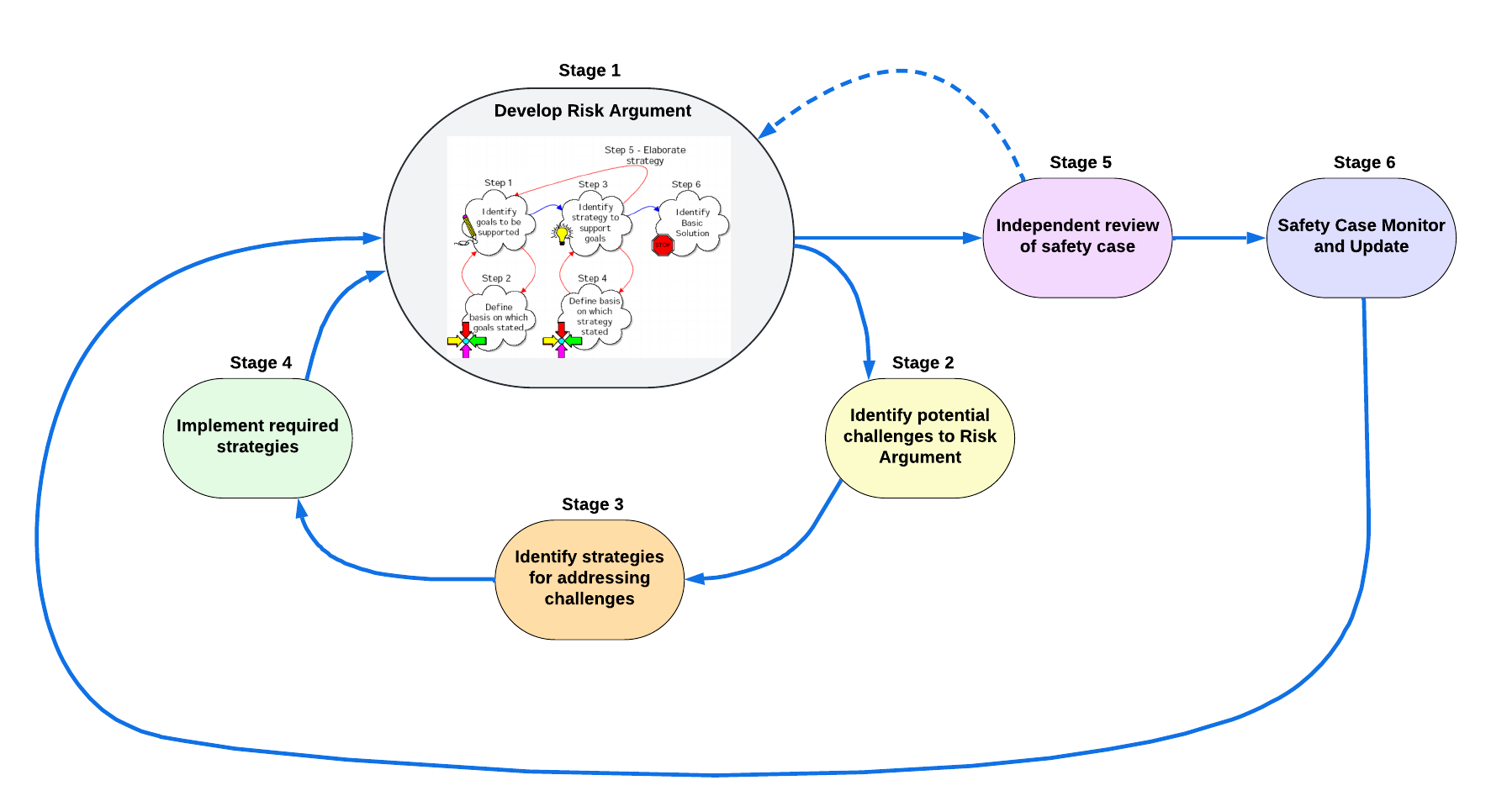}
\caption{An overview of the new method for safety case development.}
\label{fig:method}
\end{figure*}

In this section we describe a method for developing safety cases that helps to address the criticisms made against current safety case practice as discussed in Section \ref{sec:intro}. This method builds on existing good practice for safety case development, by extending the 6-step method and enhancing it with steps that explicitly incorporate review, challenge, and the provision of confidence and operational arguments as part of an iterative safety case development process. Figure \ref{fig:method} provides an overview of the method that consists of 6 stages. Stage 1 develops the risk argument, which forms the core of the safety case by structuring claims relating to the mitigation of direct causes of system hazards. Throughout the development of this risk argument Stages 2 to 4 are enacted to continually assess whether challenges to any aspect of that risk argument may exist \textit{\textbf{either now or in the future}}, and ensure that these challenges are systematically addressed in the safety case. Note that Stages 2 to 4 ensure that potentially challenges to the argument are continually and explicitly considered throughout the creation of the safety case, even if no additions or changes to the risk argument turn out to be required. Stage 5 of the method is applied once the application of the first four stages has resulted in a safety case that is ready for independent review. This independent review of the safety case may occur at multiple points throughout the safety case development process. Stage 6 considers when and how the safety case may need to be update in response to changes to the system or its operation. In this section we describe each of the stages in detail and illustrate their application using an example. Firstly we discuss the different types of argument that would be expected as part of a compelling safety case.

\subsection{Types of Safety Argument}

In order to provide a compelling, comprehensible and valid case that a system is safe for a given application in a given environment, a safety case must consist of a structured argument supported by a rigorous body of evidence that demonstrates that all of the hazards of the system have been identified and that the risk associated with each hazard has been sufficiently mitigated. Safety cases will include arguments and evidence of risk mitigation, but will also include arguments and evidence that provide confidence in those arguments of risk mitigation (such as arguments about process sufficiency or staff competency). We can refer to these different types of argument as \textbf{\textit{risk arguments}} and \textbf{\textit{confidence arguments}} respectively. Whilst both of these types of argument are essential in a compelling safety case, as discussed in \cite{hawkins2011new} it is most common to find safety cases that provide a single, unified safety argument that doesn't distinguish between these two types of argument. Presenting both types of argument in such an intermingled manner typically results in a large (often rambling) safety arguments that make it hard for the reader to grasp the core argument of risk mitigation. Thinking back to Haddon-Cave's comments in the Nimrod Review, this lack of distinction in the safety case and putting potentially relevant information into the safety case ``just in case'', leads to the lack of focus on risk that Haddon-Cave observed. It is important therefore to understand this distinction between risk and confidence arguments and to ensure the creation of the argument maintains focus on risk. Maintaining this separation is inherent within our method.

\begin{figure*}
\centering
\includegraphics[width=0.5\textwidth]{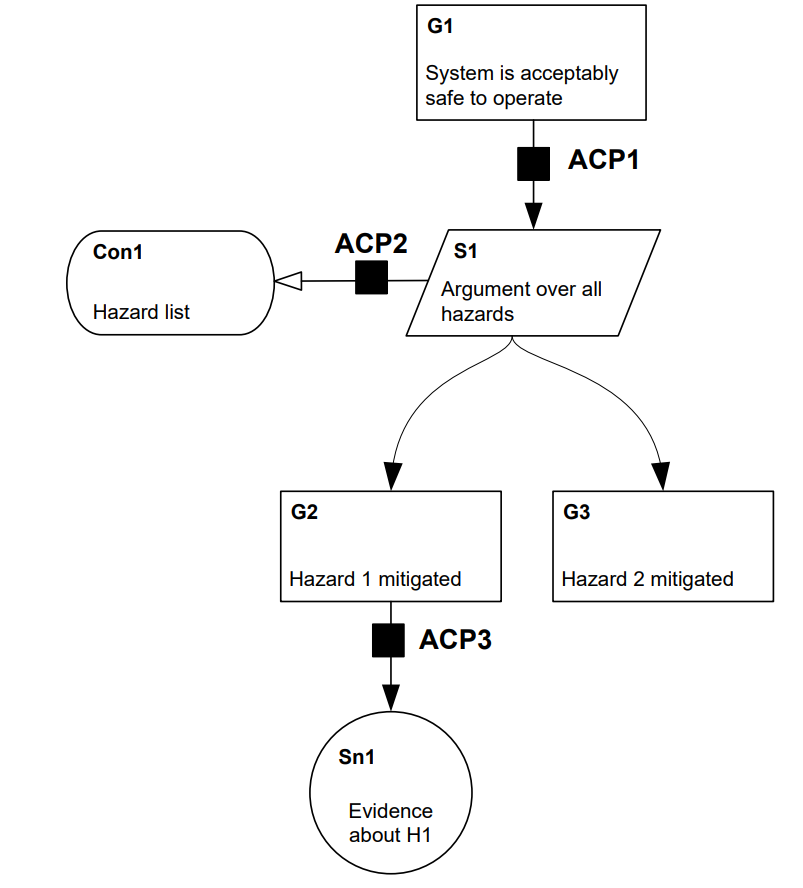}
\caption{A simple risk argument in GSN showing ACPs to link to confidence arguments.}
\label{fig:acp}
\end{figure*}

\begin{figure*}
\centering
\includegraphics[width=1\textwidth]{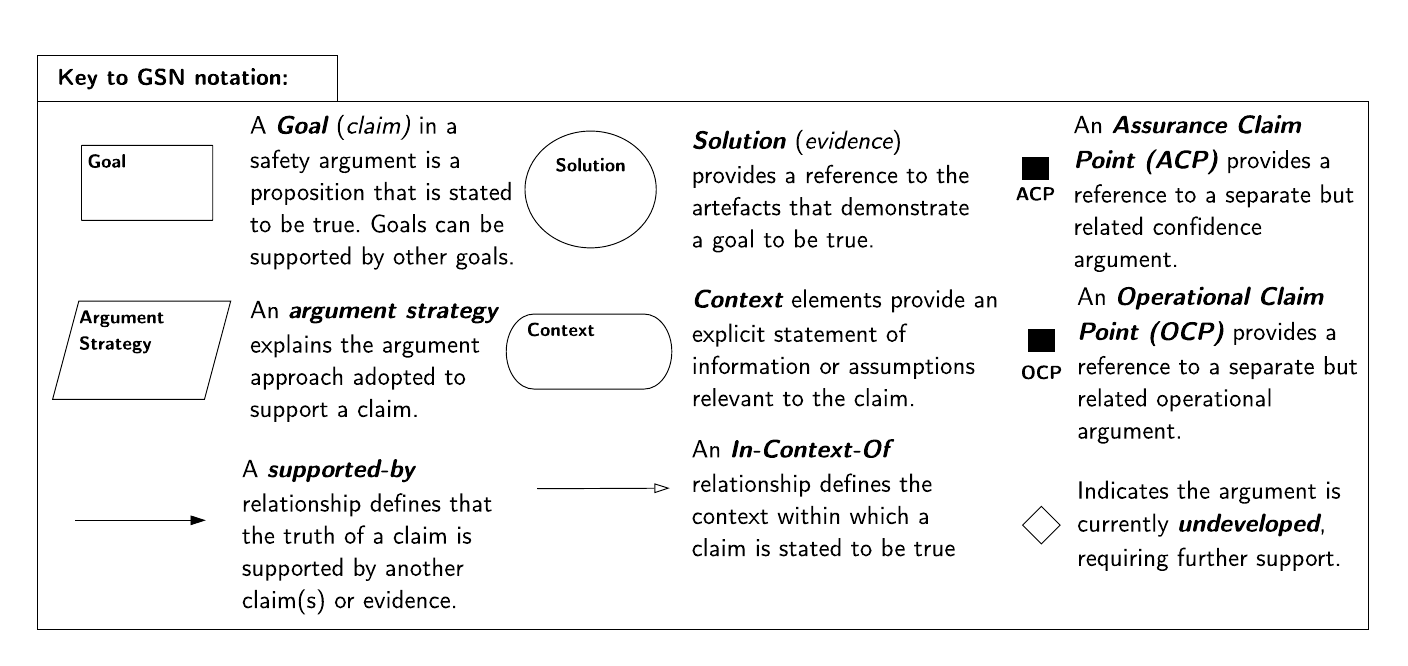}
\caption{A key to GSN notation}
\label{fig:GSNkey}
\end{figure*}

 Figure \ref{fig:acp} shows a simple example of a risk argument represented using GSN. A key to the GSN notation is provided in figure \ref{fig:GSNkey}, and full details of the notation can be found in \cite{group2021a}. Everything that is included as part of a risk argument must have a direct role as part of the causal chain to the hazard. The safety case must separately consider confidence arguments which record the reasons for having confidence in the risk argument. To ensure the focus of the safety case, confidence arguments should only be included to address specific elements within the risk argument where confidence must be demonstrated (we'll describe how to identify the need for a confidence argument later when discussing our method).
 
 Assurance claim points (ACPs) were introduced to GSN as a way to capture the relationships between the risk arguments and the confidence arguments. For example in figure \ref{fig:acp} ACP 3 indicates that there is a confidence argument demonstrating the sufficiency of the evidence about H1 (for example the integrity of the processes and tools used to generate the evidence).

Risk and confidence arguments together can provide as safety case for a system that focuses on the design and development of that system. Safety cases must also consider the safety of the operational aspects of that system (something that has been identified as crucial from the very foundation of the safety case concept \cite{cullen1993public}). As for the design elements of a safety case, current practice for operational safety cases is also often poor, with limited guidance in literature or standards on how to develop effective operational safety cases. In \cite{fenn2024new}, operational safety arguments were identified as a third type of argument that is required as part of a safety case. The ACP concept was extended to consider operational claim points (OCP) that similarly allow separate but inter-connected arguments to be created. Using OCPs enables the operational arguments to be explicitly connected to the
design-time risk arguments, ensuring clarity in both the risk argument and operational argument, whilst still ensuring an explicitly defined relationship exists. Crucially this allows design-time safety engineers to specify, in the risk argument, safety
claims relating to system operation. System operators are then able to focus on just these operational aspects of the safety case that are relevant to them.

The advantage of splitting the safety case into separate types of argument (risk, confidence and operational) is that it improves clarity by simplifying the core risk argument and ensures that the role of all claims in the argument is clear. Careful attention to linking the separate arguments provides a mechanism for guiding analysis
of the interrelationship between risk, confidence and operational arguments without overly
complex and bloated safety arguments.

\subsection{Stage 1. Develop the risk argument}

\begin{figure*}
\centering
\includegraphics[width=0.6\textwidth]{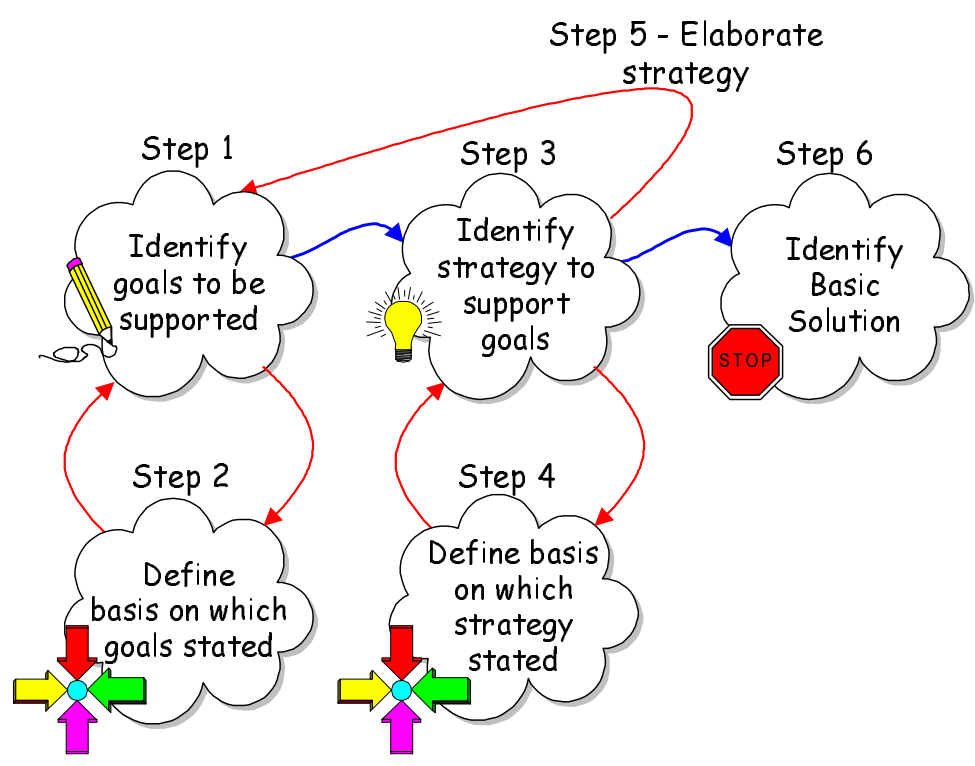}
\caption{Kelly's 6-step method \cite{kelly1999arguing}}
\label{fig:6step}
\end{figure*}

As discussed above, the core of the safety case should be an explicit \textbf{\textit{risk argument}} that demonstrates that all of the system hazards have been identified and that the risk associated with each hazard has been sufficiently managed. A systematic method should be used to create a compelling risk argument. As existing best practice for constructing structured arguments, Kelly's 6-step method \cite{kelly1999arguing} can be followed in order to create the risk argument (see figure \ref{fig:6step}). In this paper we do not repeat the existing guidance on this method, however a brief summary of the six steps is provided below.

\begin{itemize}
    \item Step 1: Identifying claims to be supported - The safety claim requiring to be supported is stated in the form of a proposition.
    \item Step 2: Define the basis on which the claims are stated - The context of the claim is examined and explicitly clarified if necessary.
    \item Step 3: Identify the strategy to support the claim - The argument strategy that will be used to support the claim is determined.
    \item Step 4: Define the basis on which the strategy is stated - The context of the strategy is examined and clarified, including any assumptions or justifications that are required.
    \item Step 5: Elaborate the strategy - Put forward the argument that elaborates the identified strategy. This will involve identifying new claims (hence returning to step 1). Steps 1 to 5 are repeated until all claims can be supported by a direct reference to evidence.
    \item Step 6: Identify basic solution - Provide explicit reference to the evidence that supports the claim.
\end{itemize}

Kelly's 6-step method reflects the hierarchical and iterative nature of safety argument construction: each cycle of steps 1 to 5 create a level of decomposition in the argument, showing how a safety claim (within a defined context) is supported through sub-claims. Kelly describes the continuing application of steps 1 to 5 until the point is reached where evidence artefacts can support a claim, at which point the development of that part of the argument terminates with the application of step 6. Stage 1 of our approach utilises Kelly's 6-step method for the first stage of the process in order to help create the risk argument. It should be noted however that Stage 1 requires the safety case developer to pause after each cycle of Kelly's 6-step method, proceeding through stages 2, 3 and 4, before returning to stage 1 to move to the next cycle of the 6-step method and so on (as shown in figure \ref{fig:method}). This stepwise iteration is a crucial aspect of our approach, since the intention is \textbf{\textit{not}} to assume that the claims in the argument are true, but to determine, through application of our method, whether this claim can be substantiated. In effect, the output of stage 1 is a proposal for how a risk claim may be supported, which will be challenged and updated in subsequent steps. Each of these iterative stages is explained in later sections.

At stage 1, the 6-step method is specifically being used to develop the risk argument. This means that the rules for risk arguments, as discussed above should be observed. Specifically that everything cited in the risk argument should have a direct role as part of the causal chain to the hazard and all claims in the risk argument must be claims about the system properties, or properties of parts of the system. An example of applying stage 1 to start to create the risk argument is described in the next section.

\subsubsection{An example of applying Stage 1}

In order to illustrate how the new method is applied, we use a simplified example of developing the safety case for an infusion pump that is used to provide controlled doses of insulin to patients. The safety argument is presented using GSN. A key to the notation is provided in Figure \ref{fig:GSNkey} and a full description of the notation is provided in \cite{group2021a}. For Stage 1 we apply the 6-step method in order to develop the first level of decomposition of the risk argument for this system.

\begin{figure*}
\includegraphics[width=1\textwidth]{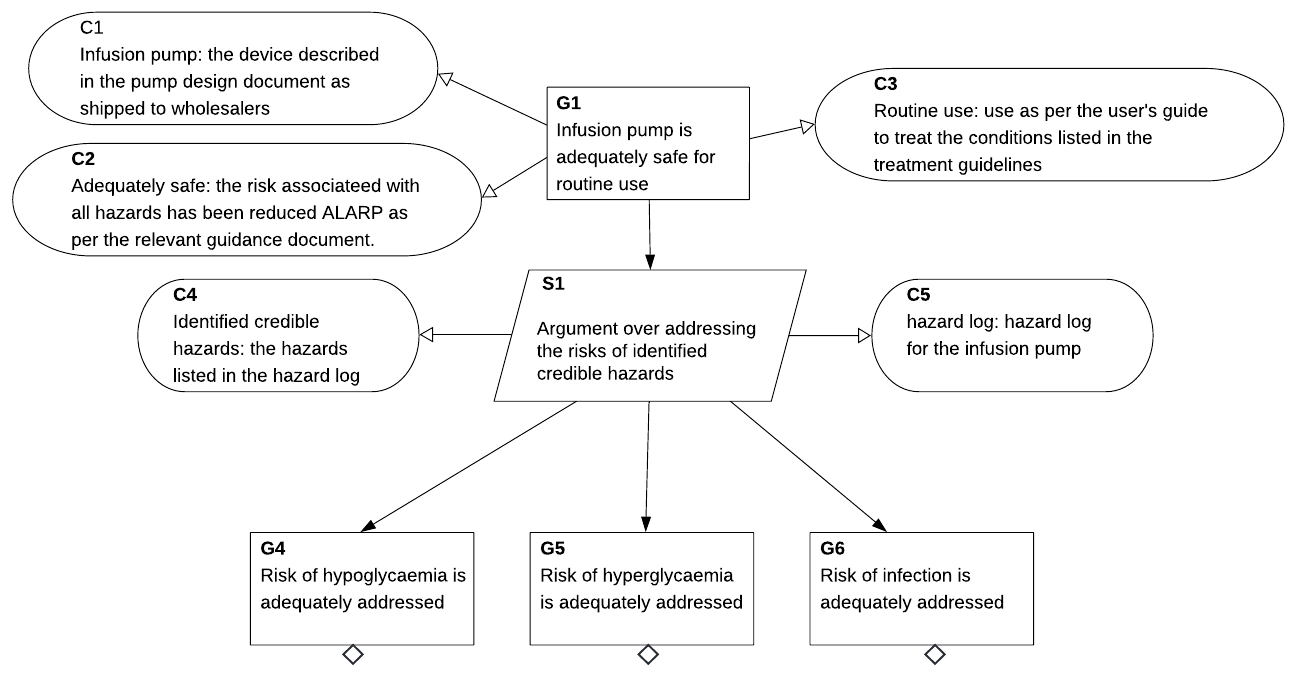}
\caption{A single level of decomposition of the risk argument for the infusion pump safety case.}
\label{fig:pump1}
\end{figure*}

\textbf{\textit{Step 1:}} The top-level safety claim is specified as shown in Figure \ref{fig:pump1} as ``Infusion pump is adequately safe for routine use'', being specific about the scope of the claim (routine use) as well as providing qualification to the claim (adequately safe). In stating this claim, we are not assuming it is true for the infusion pump, but setting out to assess whether or not that conclusion holds.

\textbf{\textit{Step 2:}} There are a number of terms expressed in the top-level claim that require clarification to ensure the scope of the safety case is understood. Specifically these are:

\begin{itemize}
    \item Infusion pump - this safety case only relates to a particular type of pump. C1 defines which pump is relevant by referencing the pump design document.
    \item Routine use - this safety case only shows the safety of the pump when used in a routine manner. C3 defines what is considered to be routine use by referencing the user's guide in the treatment guidelines.
    \item Adequately safe - it is important that the safety case is clear what is meant by ``adequately safe'' in the context of the infusion pump. C2 defines adequately safe to be that all hazards are shown to be reduced ALARP (as low as reasonably practicable) as mandated by the relevant guidance.
\end{itemize}

\textbf{\textit{Step 3:}} Guided by the definition of acceptably safe that is being adopted for the pump, the strategy for the risk argument is to support the top-level claim by addressing the risk of each of the identified hazards for the pump (to demonstrate each is reduced ALARP).

\textbf{\textit{Step 4:}} The strategy introduces ``identified credible hazards'' to the argument for the first time. It is important to make clear what these hazards are for the infusion pump. C4 specifies these to be the hazards that are listed in the hazard log. C5 references this hazard log where the up to date list of hazards is recorded. 

\textbf{\textit{Step 5:}} To elaborate the risk argument strategy that was chosen at step 3, a claim is created regarding each of the identified hazards. In this example the hazard log contains 3 hazards:
\begin{itemize}
    \item \textit{Hypoglycaemia:} Patient's blood sugar level drops below the safe minimum.
    \item \textit{Hyperglycaemia:} Patient's blood sugar level rises above the safe maximum.
    \item \textit{Infection:} Patient gets infected as a result of using the infusion pump.
\end{itemize}

Thus 3 claims (G4, G5 and G6) are created in the risk argument to support the top-level claim.

With Stage 1 completed, figure \ref{fig:pump1} shows the initial proposal for the first level of decomposition of the infusion pump risk argument.

\subsection{Stage 2. Identify potential challenges to the risk argument}

Thinking back to Toulmin's argument model, rebuttals were a key element of inductive arguments, identifying when a claim might not be true. Any safety argument or its supporting evidence is open to challenge in this way \cite{kelly2007reviewing}. Challenges are in fact themselves arguments that, rather than providing support for a claim being made in a safety case, instead cast doubt on the truth of the claim. Challenges can take the form of \textbf{\textit{counter argument}} (arguments that challenge the truth of the claim that is made) or \textbf{\textit{counter evidence}} (evidence that indicates the claim may not be true). There are different types of challenge that can be made to an argument. The term \textbf{\textit{rebuttal}} is used to refer to reasons for denying the conclusion \cite{pollock1987defeasible}. This is the case where evidence exists which supports the opposite conclusion to that put forward by a claim. Rebuttals can be thought of as direct challenges to the conclusion itself, for example a claim that failures of a certain type are not possible for a particular system is rebutted by evidence of such a failure in the system in the past. The term \textbf{\textit{undercutting}} is used to refer to reasons which attack the connection between the support and the conclusion rather than attacking the conclusion itself \cite{pollock1987defeasible}. This is the case for example when additional information questions the validity of the argument being made. Unlike a rebuttal an undercutting challenge doesn't demonstrate that the conclusion is untrue, it instead throws doubt on whether the conclusion is supported by the argument that is provided. In Toulmin's terms, undercutting could be viewed as a challenge to the warrant.

\begin{figure*}
\includegraphics[width=0.8\textwidth]{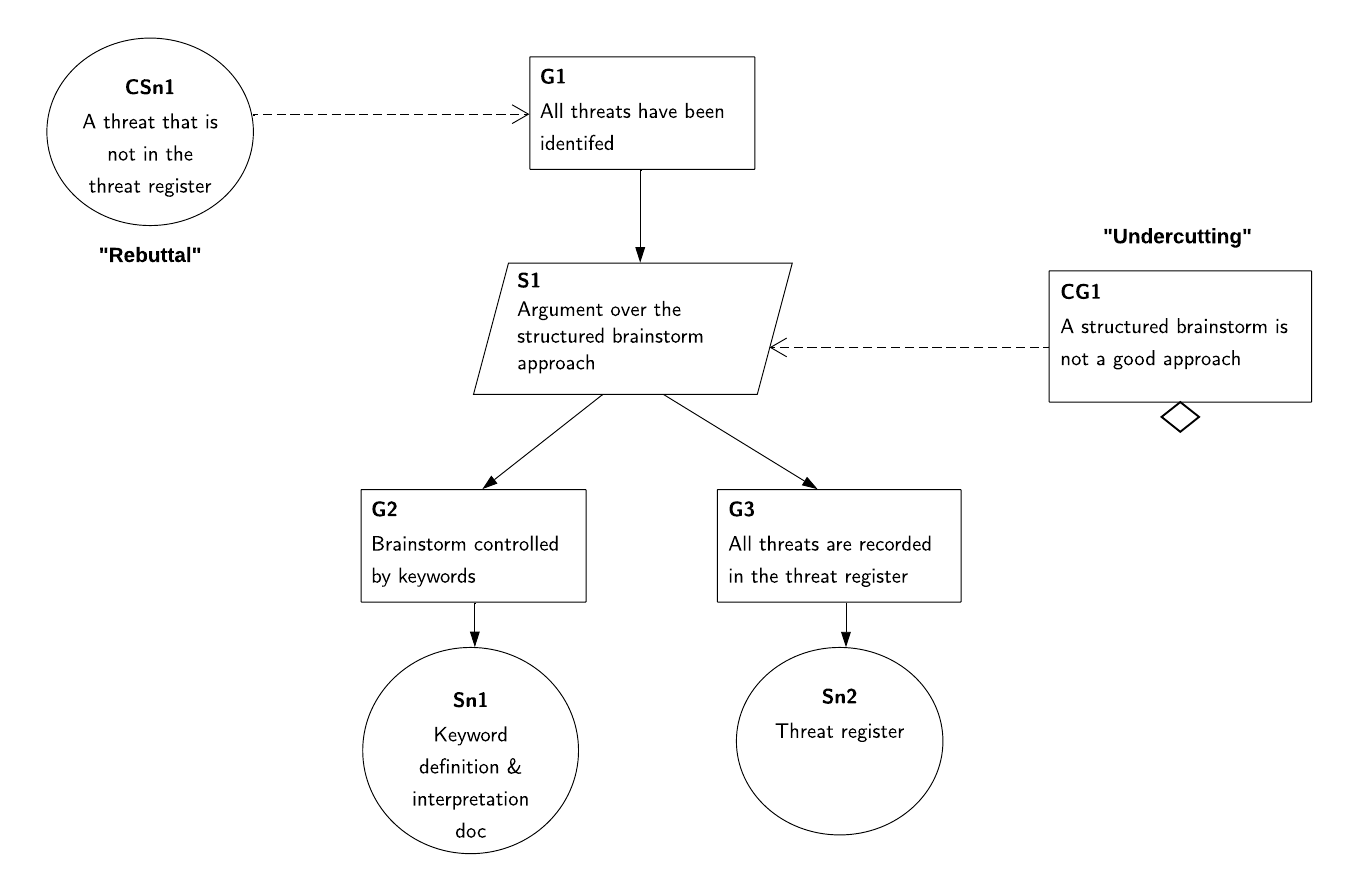}
\caption{An example of representing dialectic arguments using GSN.}
\label{fig:dia1}
\end{figure*}

\begin{figure*}
\includegraphics[width=1\textwidth]{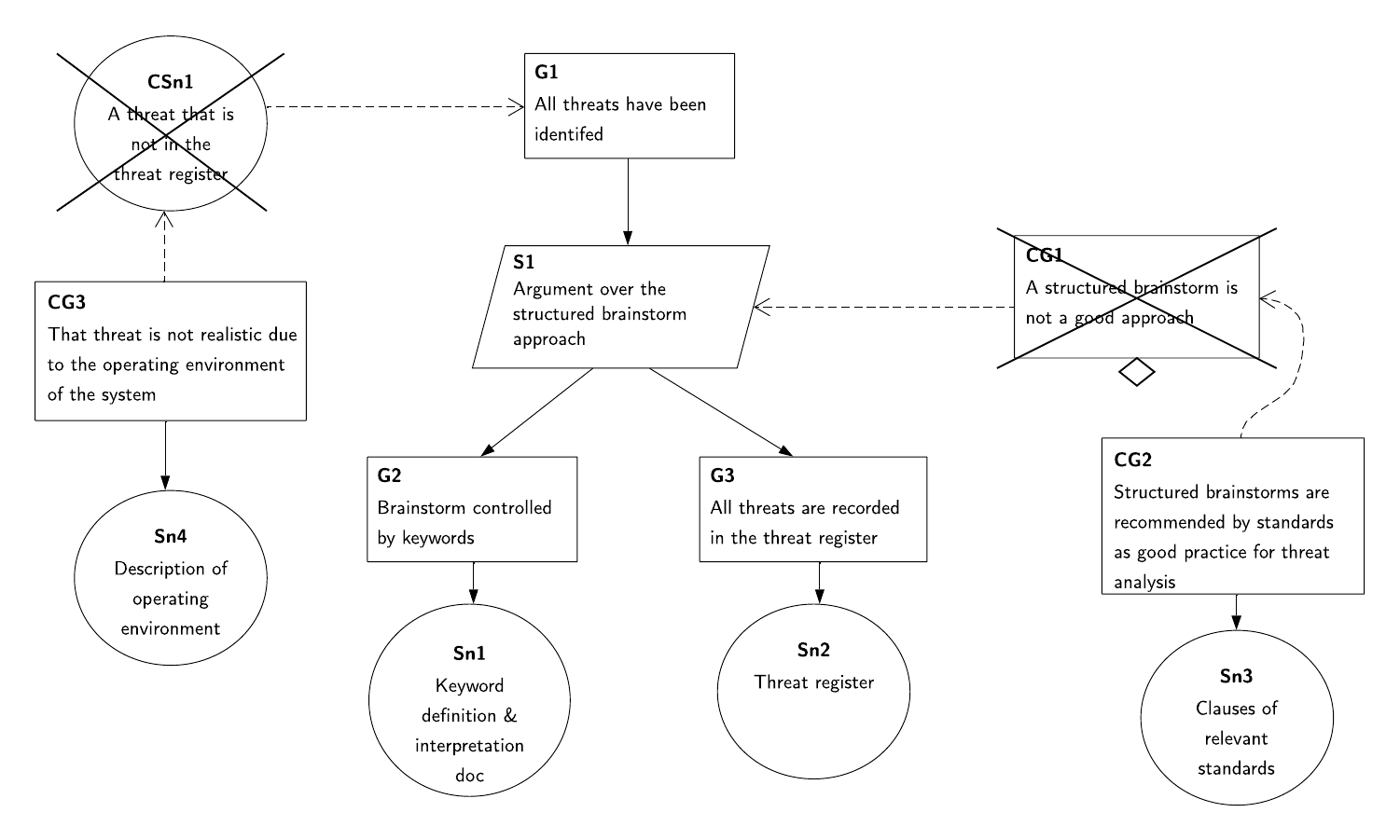}
\caption{Representing defeated challenges in GSN.}
\label{fig:dia2}
\end{figure*}

Such challenges could be made to any element of an argument; the claims, evidence, assumptions, justifications etc. could be all challenged. Valid challenges are said to \textbf{\textit{defeat}} the argument element that they challenge, which is to say that the challenge renders that aspect of the argument untrue. Not all challenges will be valid however. There may be reasons why a challenge itself is untrue (or irrelevant to the argument in the safety case), indeed all challenges are themselves open to challenge and defeat. There is an inherently dynamic nature to this. A challenge may be assumed to be valid at some point in time, casting doubt upon the truth of a safety claim, but then further information becomes available which defeats the challenge, thus enabling the truth of the claim to be re-asserted. This is one of the reasons why it is important that the development of a safety case is an ongoing, iterative process throughout the system lifecycle, continuously assessing and challenging the validity of the arguments. In this way the process of safety case challenge effectively becomes a ``conversation'' between the safety case developer and the challenger \cite{kaas2024assuring}. 
This process of logical debate between two opposing viewpoints is known as a \textbf{\textit{dialectic}} approach \cite{popper1940dialectic} \cite{finocchiaro2003dialectics}. A dialectic approach can be used in safety case development to create a more compelling argument \cite{group2021b}. GSN provides a means to explicilty represent these conversations of challenges and counter-challenges so that they can be documented and scrutinised \cite{group2021a}. Figure \ref{fig:dia1} shows an example of a rebuttal and undercutting to an argument, represented in GSN by the dotted arrows indicated a challenge relationship showing the argument element that is being challenged. Figure \ref{fig:dia2} then shows how using a dialectic approach these these challenges can themselves be challenged. In this case both the original challenges are defeated (shown not to be valid) as indicated by the crosses on those elements. Other work has also explored the idea of building a dialectic approach into safety case development. In particular the work on eliminative argumentation has shown how confidence in an argument can be increased through the identification and mitigation of argument ``defeaters'' \cite{goodenough2015eliminative} \cite{diemert2020eliminative}.

An effective safety case development process must actively seek throughout the system lifecycle to identify and, where possible address, safety case challenges. Current practice is generally to treat the identification of challenges as a reviewing activity that occurs once the development of the safety case is completed. This would be expected to be performed primarily or solely by someone other than the safety case developer. There are a number of reasons why this is undesirable:
\begin{itemize}
    \item Developers are not then encouraged to think about ways in which the safety case may be invalidated, but rather to focus only on the positive aspects of the safety case potentially encouraging a mindset of confirmation bias.
    \item It encourages the development of the safety case to simply reflect the system as already designed, rather than challenging design decisions as the development of the system progresses, potentially encouraging an after-the-fact assurance approach.
    \end{itemize}

Instead, the method described in this paper requires the systematic consideration of challenges as an integral part of the safety case development. At each decomposition of the risk argument it is required that potential challenges to the argument are identified. These challenges should be identified by the safety case developers themselves (the developer also taking on the role of challenger). This enforces a discipline of continuous self-review throughout argument development.

The most important part of this stage is that the we are trying to identify the significant \textbf{\textit{potential}} challenges to the risk argument. Particularly early on in the development process, evidence to support a potential challenge may not exist, however if it is a realistic challenge to the validity of the risk argument that may be supported at some later point, that should be identified at this stage. It may later be revealed that such a challenge cannot be supported, in which case this can be removed from the safety case. To ensure this stage is tractable, when identifying challenges it is important to be reasonable and proportionate. The focus should be on significant challenges that could genuinely threaten the validity of the safety case, rather than identifying lots of trivial challenges whose impact is low. It is also important to realise that there may not be any significant challenges identified at every level of decomposition in the risk argument. This is clearly an acceptable outcome, and the act of checking and not identified anything significant is, in itself, an important activity that builds confidence in the safety case.

\subsubsection{An example of applying Stage 2}

\begin{figure*}
\includegraphics[width=1\textwidth]{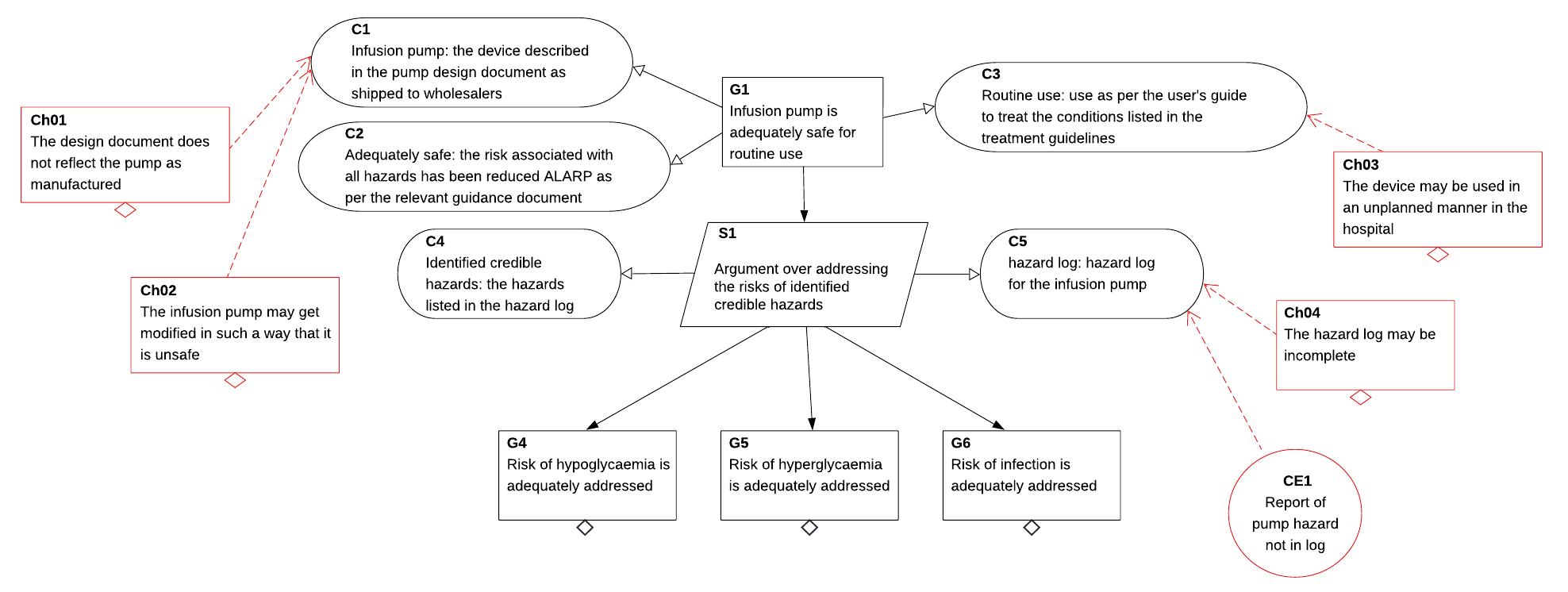}
\caption{Adding potential challenges to the infusion pump risk argument.}
\label{fig:pump2}
\end{figure*}

If we now return to our example infusion pump safety case we can systematically identify a number of potential challenges to the risk argument that was created earlier by considering each element of that argument. This lead us to identify the challenges shown in Figure \ref{fig:pump2} which are interconnected to the relevant risk argument elements using dotted lines in accordance with GSN dialectic notation. 

We discuss each of these challenges in turn.

\begin{itemize}
    \item \textbf{\textit{Ch01:}} The design document that is used to describe the infusion pump may not be a true reflection of the infusion pump as it is manufactured. If the design document is out of date, or contains errors that make it inconsistent with the pump this challenges context C1, so the top-level claim may not be true.
    \item \textbf{\textit{Ch02:}} If the user of the pump makes any modifications to the pump then it may no longer correspond to what is described in the design document, invalidating the context C1.
    \item \textbf{\textit{Ch03:}} When the pump is actually being used, the way in which it is used may not correspond to what the user's guide describes, challenging context C3.
    \item \textbf{\textit{Ch04:}} The hazard log may be incomplete due to insufficiencies in the hazard identification process. If there are hazards associated with use of the infusion pump that are not in the hazard log these will not be addressed in the safety case.
    \item \textbf{\textit{CE1:}} It is identified from previous operation of infusion pumps, that there is a hazard relating to allergic reactions to the materials that the pump is made from. This hazard is not currently included in the hazard log. This indicates that the hazard log (and hence the argument) may be incomplete. 
\end{itemize}

\subsection{Stage 3. Identify strategies for addressing the challenges}

For each of the potential challenges identified at step 2, the way in which these challenges can be addressed must be determined. The way in which potential challenges are dealt with will depend upon the nature of the challenge. There are essentially five options that can be considered: 

\begin{enumerate}
    \item \textbf{\textit{Rebuttal:}} Rebut the challenge.
    \item \textbf{\textit{Change:}} Make changes to the system or the risk argument in order to address the challenge.
    \item \textbf{\textit{Confidence argument:}} Provide a confidence argument that addresses the challenge. 
    \item \textbf{\textit{Operational argument:}} Provide an operational argument that addresses the challenge.
    \item \textbf{\textit{Monitoring:}} Identify a requirement for monitoring.
\end{enumerate}

We discuss each of these options in turn.

\textbf{Option 1: \textit{Rebuttal}}

As discussed previously, not all challenges that may be identified within a safety case are valid. As with any claims in a safety case, challenges are asserted propositions which are therefore subjective. Challenges may themselves be challenged in the same way that elements of the argument itself can be challenged (as seen earlier in figure \ref{fig:dia2}). These challenges normally take the form of rebuttals which provide argument and evidence that demonstrated the claim is untrue. For this stage of our process, the safety case developer can rebut a proposed challenge on the grounds that its an invalid challenge in the context of the safety argument being presented, i.e. this challenge is infeasible given what is known about the system and its operation. This may be the developer revealing why a potential challenge they themselves have identified to their own argument is not valid. It is good practice for the developer to explicitly document these challenges and rebuttals as part of the safety case. Showing that such challenges have been identified and addressed increases confidence in the safety case by demonstrating that possible defeat of the safety argument has been systematically considered.

\textbf{Option 2: \textbf{\textit{Change}}} 

Where challenges are assessed as being valid, they may represent a challenge to the risk argument itself i.e. that the argument presented is not a valid strategy for the system under consideration. An alternative structure for the risk argument would therefore be required. For example, the risk argument may present a particular mitigation strategy for a hazard that is shown through a challenge to be unviable to implement for this system. In such a circumstance an alternative mitigation strategy must be identified, and the risk argument updated to reflect this.
Although such a design modification \textit{may} have emerged from the development process, systematically assessing the safety case in this way can increase confidence, early in the system lifecycle, in the appropriateness of the chosen risk mitigation strategies.

\textbf{Option 3: {\textit{Confidence argument}}} 

Some challenges will relate to potential uncertainties in the risk argument. These do not represent direct rebuttals of the risk argument, but instead relate to issues that could potentially undermine confidence in the argument element. For example a challenge may relate to uncertainty in whether an artefact that provides context to an argument is correct and up to date. To address this challenge requires that confidence be established in the correctness of the artefact, for example by considering the process used to create the artefact, or the competency of the people that created it. This confidence can be provided explicitly as part of the safety case through the provision of a confidence argument that specifically addresses the uncertainty in the artefact (through, for example, arguing over the process that was used). As discussed earlier, an ACP can be used to indicate where in the risk argument the confidence argument is provided.

\textbf{Option 4: \textit{Operational argument}} 

Some potential challenges may relate to operational aspects that could undermine the validity of the risk argument. For example a challenge may relate to an assumption about the system that will only hold true throughout operation if certain operational measures are in place. Although the challenge itself won't be supported unless and until counter-evidence arises during operation the challenge should be raised as part of this stage of the process at design-time as a potential challenge for which the safety case must consider mitigation. This enables necessary assumptions on operation which are crucial to the validity of the safety case to be explicitly identified during the creation of the safety case. For example, claims regarding the failure rate of a component made in the system risk argument may be based on analysis that relies on assumptions about the frequency with which maintenance of the component is undertaken. This part of the risk argument would therefore be challenged if this maintenance assumption turned out to be untrue. To address this challenge requires an operational argument to demonstrate that maintenance is being undertaken as required by the risk argument. Although this operational argument will not be fully supported until the system is operational (and evidence of correct maintenance can be collected), the need for such an argument should be explicitly identified at this stage and captured in the risk argument \cite{fenn2024new}. As discussed above, an OCP can be used to indicate where in the risk argument an operational claim is required.

\textbf{Option 5: \textit{Monitoring}} 

Similar to option 4, there may also be potential operational challenges identified that relate to the possibility of counter-evidence to the risk argument arising during operation. For example failure modes may be observed in a component at run-time that had not been identified as part of the analysis undertaken during the development of the system. Although at this stage of our safety case process we cannot know whether such counter-evidence exists (as the system has not yet been deployed), since it is known that its presence will challenge the risk argument, the necessity of implementing run-time monitoring to identify the occurrence of the counter-evidence should be considered. For example a failure reporting system could be introduced into the system that would identify and record failures of components during operation. Such monitoring requirements help to mitigate system risk, but effective monitors can only be put in place if it is known what must be monitored. Through identifying the potential run-time challenges to the risk argument during development, effective operational monitoring can be designed into the system \cite{hawkins2023identifying}.

\subsubsection{An example of applying Stage 3}

If we return to the challenges to the infusion pump risk argument that were identified at Stage 2 (Figure \ref{fig:pump2}), we can identify a strategy to best address each of those challenges within the safety case.

\begin{figure*}
\includegraphics[width=1\textwidth]{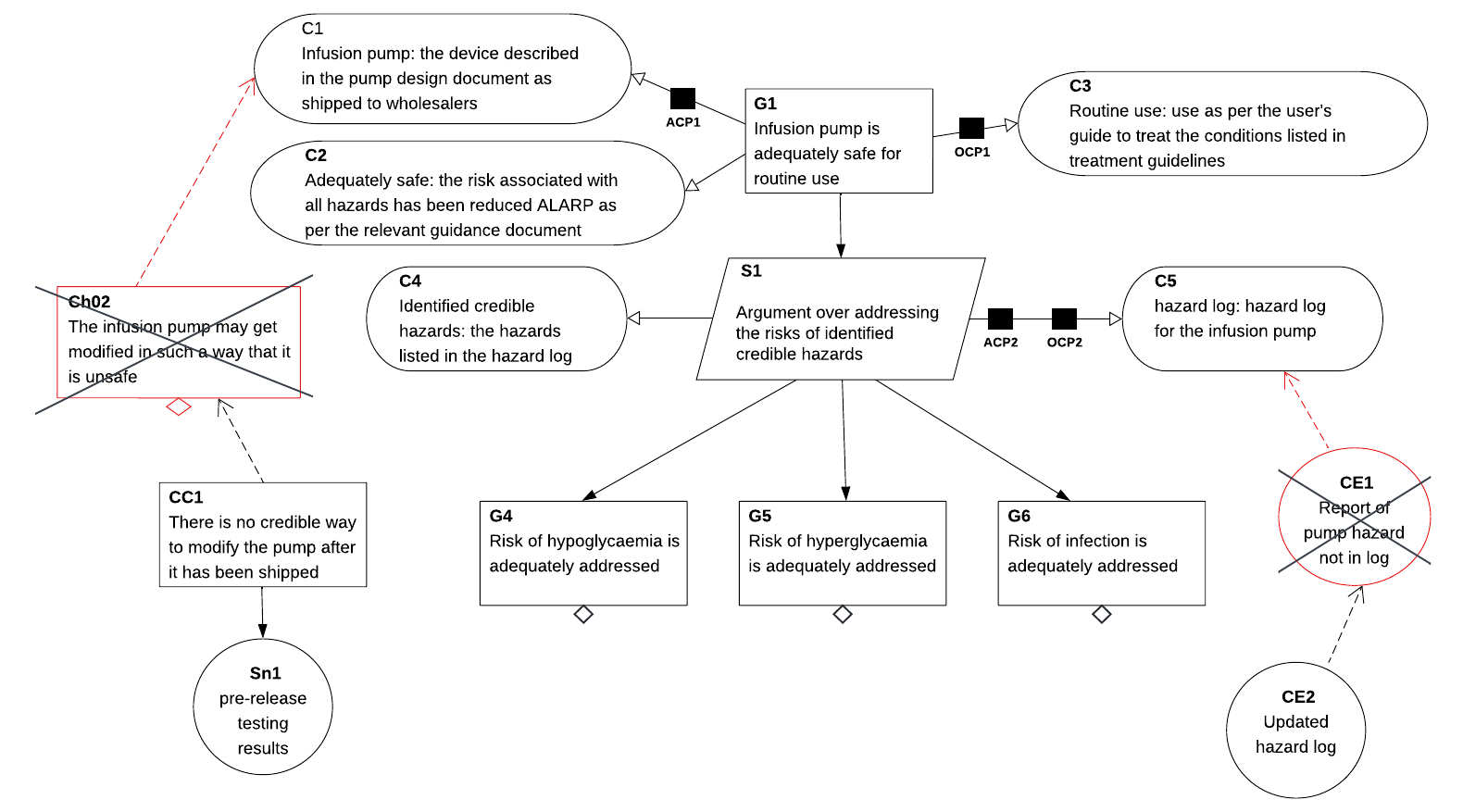}
\caption{Identifying strategies to address the potential challenges to the infusion pump risk argument.}
\label{fig:pump3}
\end{figure*}

We discuss each of challenge in turn:

\begin{itemize}
    \item \textbf{\textit{Ch01:}} Addressing Ch01 (that the design document may not be a true reflection of the pump) requires that confidence is demonstrated, as part of the safety case, that the pump design document is a true representation of the pump as manufactured. This can be achieved through the creation of a confidence argument specifically addressing this. This confidence argument will be provided at ACP1 (as shown in Figure \ref{fig:pump3}).
    \item \textbf{\textit{Ch02:}} The development team are able to confirm that due to the design of the pump, they do not believe that there is a credible way in which the pump can be modified once it leaves the factory. The pre-release testing that they undertake supports this by demonstrating that the pumps are tamper-proof. This provides a rebuttal which defeats this challenge (as shown in Figure \ref{fig:pump3}).
    \item \textbf{\textit{Ch03:}} Addressing Ch03 (that the device may be used in an unplanned manner) requires that that it can be demonstrated that people using the pump are following the guidelines in the user's guide. This is something that requires consideration of the operational aspects of the pump, so should be addressed as part of the operation safety case. The creation of OCP1 in Figure \ref{fig:pump3} ensures that there is an explicit link from the risk argument to the relevant operational argument.
    \item \textbf{\textit{Ch04:}} In order to fully address challenge Ch04 (the completeness of the hazard log) there are two distinct strategies that can be adopted. Firstly, confidence that the list of hazards in the hazard log is complete must be demonstrated through a confidence argument that considers the process by which the hazard log was created. This confidence argument will be provided at ACP2. In addition, since the completeness of the hazard log is crucial to the validity of the argument, we can also monitor for the occurrence of any new hazards that only become apparent once the pump is in use. This will require a process for monitoring and reporting hazards that the users of the system can follow. This must be included in the operational safety case, and is captured by OCP2 in Figure \ref{fig:pump3}.
    \item \textbf{\textit{CE1:}} Action is required to address the additional hazard of allergic reaction. The immediate response is to rebutt the counter evidence by updating the hazard log to include this hazard (as shown in Figure \ref{fig:pump3}). Further measures in response to the counter evidence are discussed at step 4.
\end{itemize}

\subsection{Stage 4. Implement required strategies}

Applying stages 1 to 3 of our method as discussed above ensures that the safety case developer has identified the argument strategies that are required in the safety case in order to further develop a compelling saftey argument. This includes both the strategies to be adopted in order to further develop and decompose the risk argument (step 5 of the standard 6-step-method process), as well as the strategies chosen in Stage 3 of our approach to address the challenges identified at Stage 2. In terms of the risk argument, this could involve modifying the risk argument structure, creating the next level of argument or evidence decomposition, and adding required ACPs and OCPs to elements of the risk argument. This may then also necessitate the development of confidence or operational arguments relevant to the defined OCPs and ACPs.

Only once all of these required strategies have been implemented should the next stage of the method be enacted, which, as shown in \ref{fig:method} involves repeating stage 1. In doing so, the method ensures that due consideration is given throughout the entire development of the safety case to how it may be undermined or challenged,resulting in a more robust and defensible case. 

Stages 2 and 3 identified a number of dialectic argument elements and corresponding strategies which mitigate the challenges. Technically the dialectic elements could be removed from the safety case once the mitigations have been implemented (and the challenges addressed). There is however huge value in maintaining the history of the challenge and counter-challenge so that the safety case reader is aware that potential challenges were identified and addressed. Being able to see that the challenges were identified provides confidence to the reader that all issues are addressed and that the effects of confirmation bias have been actively mitigated. If this dialectic history is lost, it is less clear to the reader why the safety argument approach adopted in the safety case was chosen, and why it is sound.

Of course leaving all dialectic elements created throughout the method visible in the safety case will very quickly lead to a potentially overwhelming amount of information in the safety case, and more importantly this is likely to distract the reader from understanding the core risk argument, defeating one of the key advantages of a risk-based argument approach. A compromise solution is therefore to make the dialectic aspects of the argument available to the reader when required, but not to leave them represented on the core argument. With simple tool support such aspects could be easily turned on and off in the safety case as required, or could be provided as a separate `view'.

\subsubsection{An example of applying Stage 4}

\begin{figure*}
\includegraphics[width=1\textwidth]{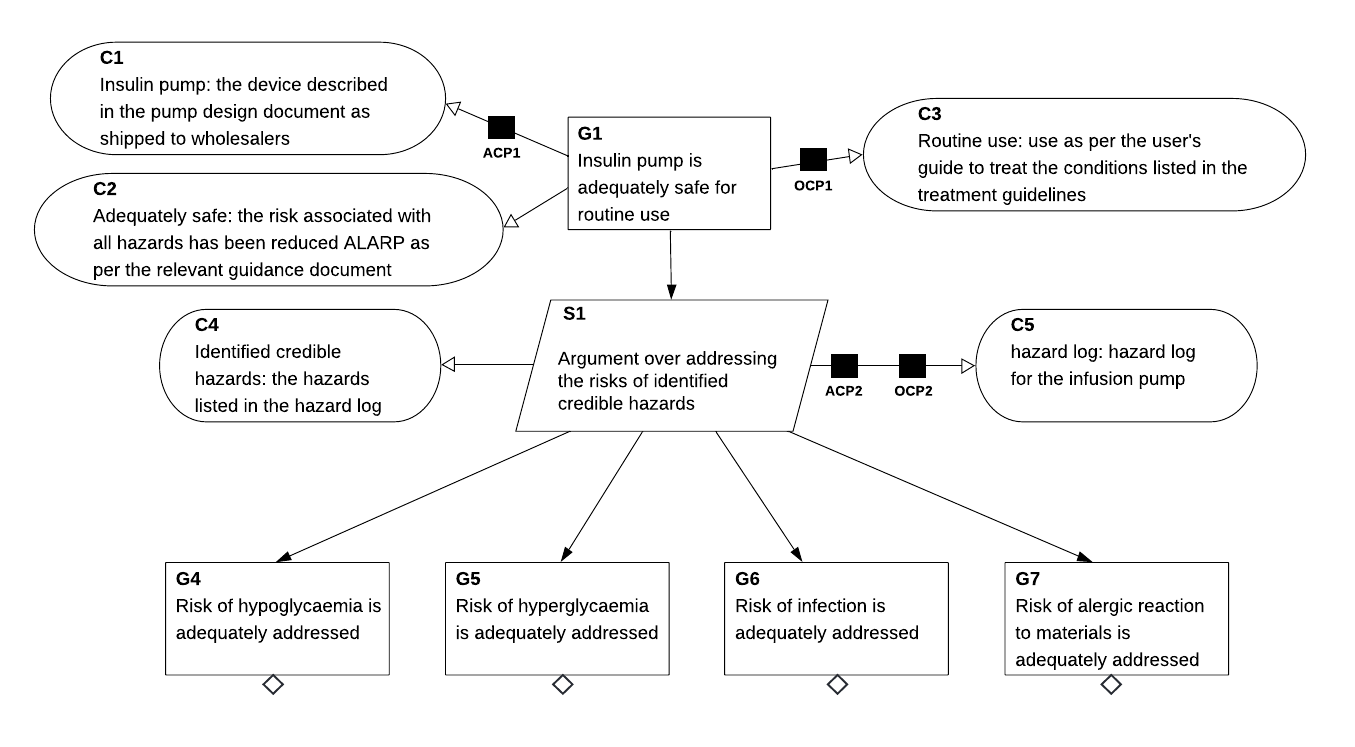}
\caption{Changing the infusion pump risk argument to address potential challenges.}
\label{fig:pump4}
\end{figure*}

For the infusion pump safety case example, the strategies that were identified at Stage 3 for addressing the potential challenges have been implemented. The resulting argument is shown in Figure \ref{fig:pump4}. Again we take each of the challenges in term and discuss how the safety case was developed using the identified strategies.

\begin{figure*}
\centering
\includegraphics[width=0.8\textwidth]{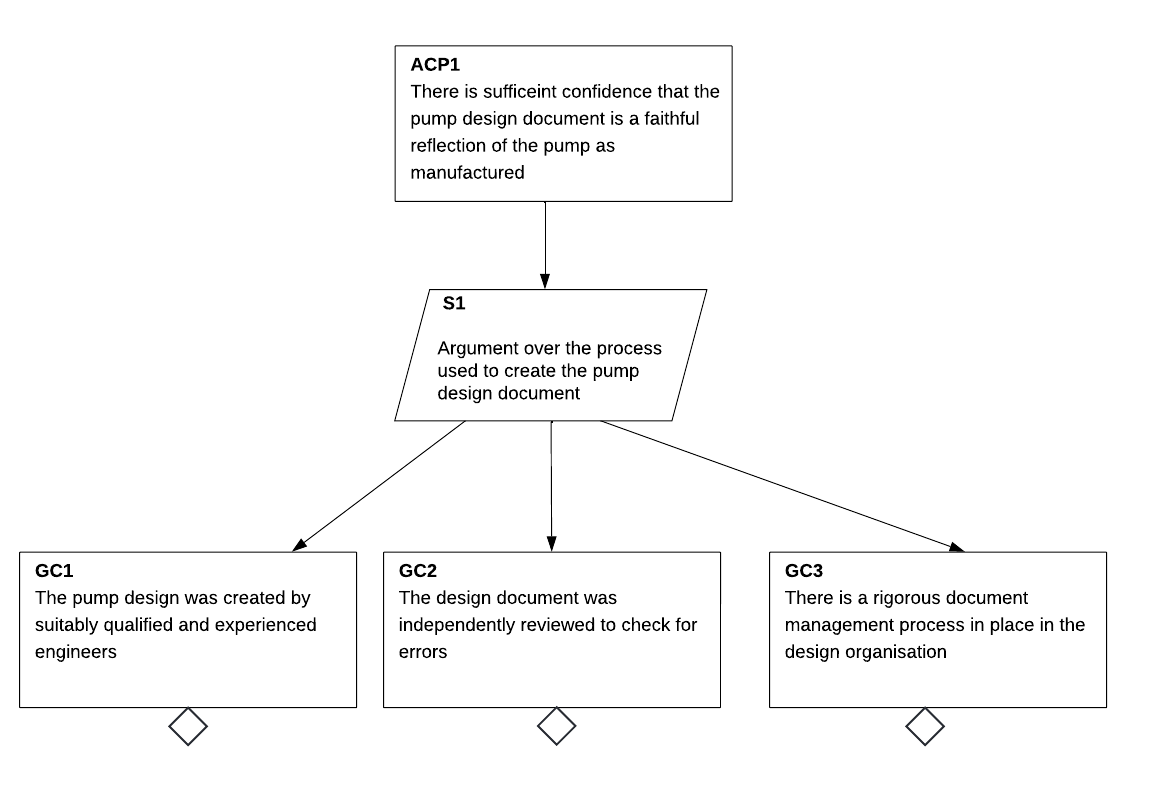}
\caption{Confidence argument for ACP1.}
\label{fig:pump5}
\end{figure*}

\begin{figure*}
\centering
\includegraphics[width=0.7\textwidth]{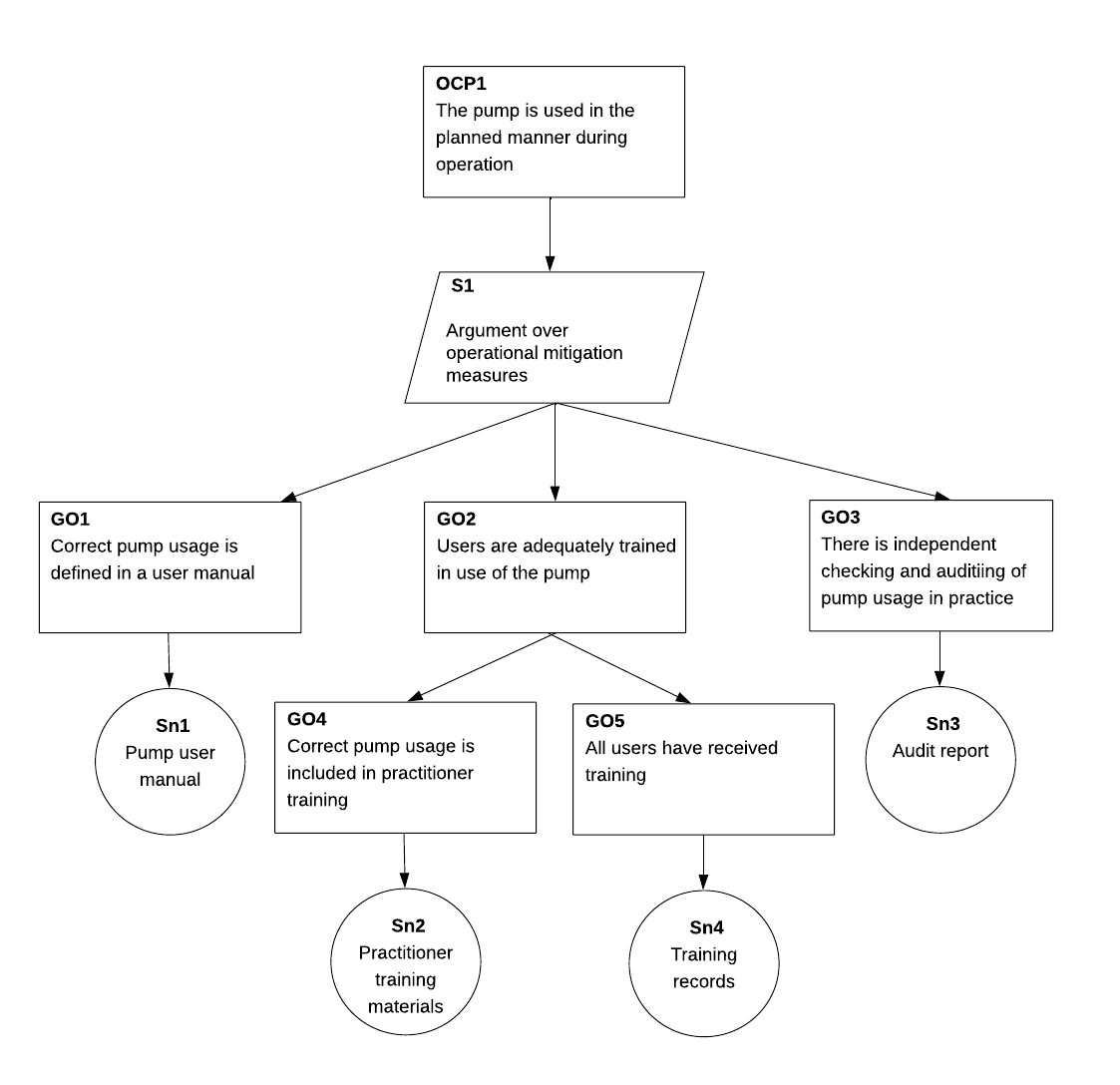}
\caption{Operational safety argument for OCP1 (part of the operational safety case for the infusion pump).}
\label{fig:pump6}
\end{figure*}

\begin{figure*}
\centering
\includegraphics[width=0.8\textwidth]{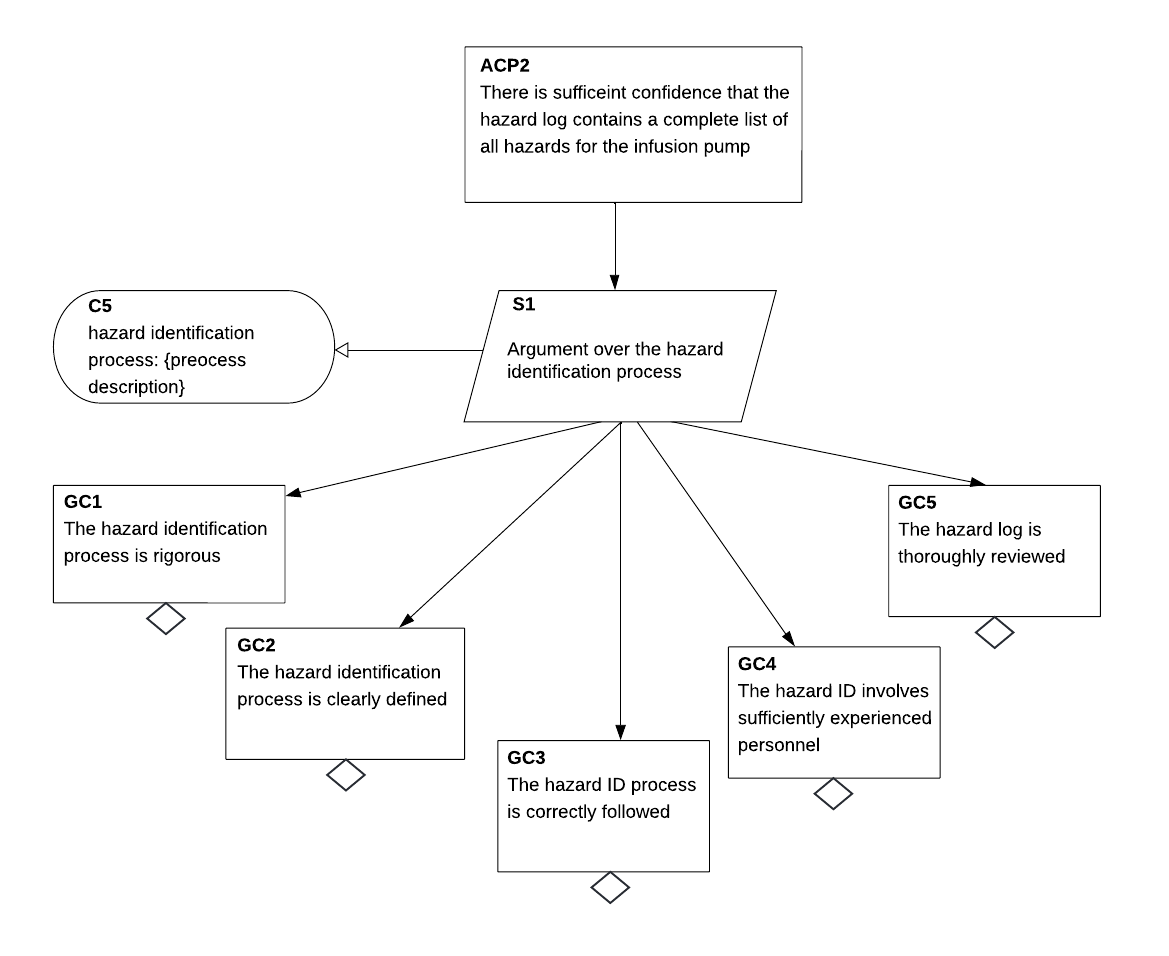}
\caption{Confidence argument for ACP2.}
\label{fig:pump7}
\end{figure*}

\begin{figure*}
\centering
\includegraphics[width=0.6\textwidth]{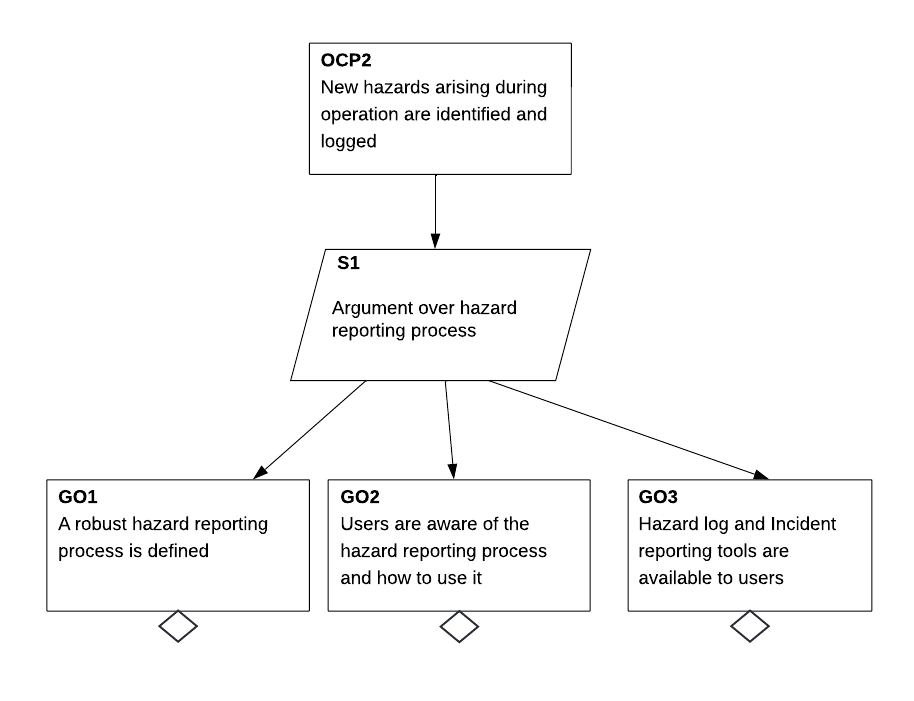}
\caption{Operational safety argument for OCP2 (part of the operational safety case for the infusion pump).}
\label{fig:pump8}
\end{figure*}

\begin{itemize}
    \item \textbf{\textit{Ch01:}} A confidence argument to address ACP1 is created. The high-level structure of this argument is shown in Figure \ref{fig:pump5}. The confidence argument addresses the claim ACP1 that \textit{``There is sufficient confidence that the pump design is a faithful reflection of the pump as manufactured''}.  This is demonstrated through consideration of the process that is used to create the design document. This includes consideration of the qualifications and experience of the engineers that created the document, the reviewing of the document to identify any errors, and the sufficiency of the document management processes that are in place in the pump design organisation. In Figure \ref{fig:pump5} we don't develop this confidence argument further, however each of these sub-claims must be supported be supported by compelling argument and evidence.
    \item \textbf{\textit{Ch02:}} Since this challenge is successfully rebutted by the testing that shows it cannot be modified, the challenge is removed from the argument in Figure \ref{fig:pump4}. It is important however that the identification and discharge of this challenge is not lost since it shows a potential issue that could have affected the strength of the safety case has been considered. A record of this dialectic aspect of the safety argument is therefore maintained, even though for clarity of the safety case we choose not to show it in the risk argument.
    \item \textbf{\textit{Ch03:}} An argument is created in the operational safety case to address OCP1. This is done by demonstrating that the pump is used correctly during operation. This argument, shown in Figure \ref{fig:pump6}, provides a number of claims relating to the use of the pump in operation. Firstly it is shown that correct usage has been defined (GO1). Secondly, it is shown that the users are adequately trained in the use of the infusion pump (GO2). Finally it is demonstrated that independent checking and auditing is taking place during operation to confirm that pumps are being used correctly (GO3). As shown in Figure \ref{fig:pump6}, all of these claims will be supported by evidence, however note that some of this evidence, such as the audit reports (Sn3) and training records (sn4), will not be available until the pumps are in use. So although this operation argument is produced at development time, some of the evidence will not be provided until during operation.
    \item \textbf{\textit{Ch04:}} Addressing Ch04 required both a confidence argument to address ACP2 and an operational argument to address OCP2. The confidence argument for ACP2 is shown in Figure \ref{fig:pump7}.The confidence argument addresses the claim that \textit{``There is sufficient confidence that the hazard log contains a complete list of hazards for the infusion pump.''}. Similarly to ACP1, this is demonstrated through consideration of the hazard identification process. This includes the rigour of the process, the fact it is defined clearly and followed by qualified people, and is subject to thorough review.
    The operational argument to address OCP2 is shown in Figure \ref{fig:pump8}, and would be included as part of the operational safety case. This argument focuses on the hazard reporting process to identify new hazards that may arise during operation. The argument shows that there is a robust hazard reoprting process in place, that the pump users are aware of this process, and that the required reporting tools are available to the users. As with OCP1, some of the evidence required to support these claims will not be available until the pumps are in use.    
    \item \textbf{\textit{CE1:}} Addressing the identification of the additional ``allergic reaction'' hazard involved updating the hazard log, however the risk argument itself must also be updated to reflect this. This can be seen in Figure \ref{fig:pump4} where G7 has been added as an additional safety claim relating to that hazard. Support for G7 will then be developed further with argument and evidence (as for G4, 5 and 6) as the process continues through further iterations of the stages.
\end{itemize}

\subsection{Stage 5. Independent review of safety case}

As shown in figure \ref{fig:method}, Stages 1 to 4 are repeated to develop each level of the risk argument, identifying challenges and their mitigations and updating the argument to implement those mitigations. This cycle continues alongside the system development in order to build up the safety case until, ultimately, all claims are supported by evidence. As we have seen, inherent within the new method is an aspect of self-review (at Stage 2) where the safety case developers themselves seek to find potential challenges to their argument. Also crucial to an effective safety case process is \textit{independent} review.

Independent review is review of the safety case by people not actively involved in the development of the safety case. That could be both people within the developing organisation who have not been involved in development of the safety case, or people who are external to the developing organisation (such as auditors, regulators, customers etc.). The purpose of the independent review is essentially the same as for the developers at stage 2, to try to identify potential challenges to the argument and evidence presented in the safety case that may have been overlooked by the safety case developers. In particular independent review can be invaluable in bringing in knowledge of relevant counter-evidence from elsewhere, as well as identifying undocumented aspects of the safety case, such as unstated assumptions or implicit and unsubstantiated argument.

It can be seen in figure \ref{fig:method} that independent review is undertaken after safety case development stages 1 to 4 are completed. This does not mean that independent review only happens once the safety case is "complete"; it may be appropriate to undertake independent review at any point in the safety case development process at the end of any development cycle (after stage 4). Often good practice is to align independent safety case review with key stages or milestones of the system development process.

The iterative nature of the first 4 stages of the safety case method should ensure that there is less likelihood that major problems or weaknesses with the safety case will be identified through independent review; indeed it may be hoped that effective safety case development using this new method should identify and address potential problems early in the safety case development process. One of the advantages of the method is that it de-risks the safety case process from a project perspective by reducing the chance that safety case problems are only identified late on, when system design and development commitments have been made and are very time consuming and expensive to fix.

Once the independent review stage is complete, the process will either return to Stage 1 to develop the safety case further (to complete an incomplete safety case or respond to findings of the independent review) or, if both development of the safety case and its review are complete, then the project would move to a monitoring and update phases (Stage 6).

Like for the internal review activities undertaken earlier in the process, the independent review can make use of a dialectic argument approach to identify and document the challenges, counter argument and evidence etc., bringing the advantages of explicit representation of these in the safety case. In any case, it is crucial at this stage that an effective and independent review occurs, and that any findings and recommendations are clearly communicated to the safety case developers and acted on effectively.

\subsection{Stage 6. Safety case monitor and update}

Even once a safety case has been developed and independently reviewed, it is imperative to continue to monitor the validity of the safety case and, where necessary, to update the safety case throughout the life of the system \cite{papadopoulos2000safety}, \cite{denney2015dynamic}. There are a number of reasons why a previously acceptable safety case may become challenged during the system lifecycle. These include:

\begin{itemize}
    \item The occurrence of counter-evidence
    \item Changes to the system or its operation
    \item Challenges revealed from through-life safety management activities
\end{itemize}

Counter evidence that challenges the safety case could arise at any point in the system lifecycle. For example during testing it may be discovered that a system component is unable to satisfy a requirement about which a safety claim has been made in the safety case. Or for example during system operation the usage of the system may change in a way that wasn't anticipated in the safety case (such as extending its operational hours). It is important that effective and active monitoring for counter-evidence such as this is in place. Without that monitoring, the counter evidence may occur, invalidating the safety case, without the system users' knowledge and therefore with no ability to respond. As part of Stage 3 it was discussed how specific requirements on monitoring for counter-evidence can be identified during safety case development. These monitoring mechanisms can then be enacted in the system itself, providing the necessary awareness of the existence of counter evidence. Also, more general mechanisms for identifying the occurrence of counter evidence should be used. This could include, for example, processes to ensure knowledge regarding operational events for other similar systems or information from component suppliers or system operators is obtained in a timely manner. Such mechanism should be included as part of the system's safety management system (SMS).

A natural trigger for the assessment of safety case challenges is when changes are made to the system due to upgrades or modifications, as well as any operational changes. Again there is a crucial role here for the SMS in defining change processes that include consideration of the impact of changes on the safety case. The SMS also provides a mechanism for implementing periodic reviews of the safety case through the system lifecycle, and is a crucial part of achieving a successful safety case regime in an organisation.

By whatever means the potential challenges to the safety case are identified, their impact must be assessed to identify if updates are required to the safety case in response. This requires identifying the elements of the safety argument that are directly affected by the challenge (claims, assumptions, evidence etc.) as well as the implications of this on the rest of the argument that this supports. In some cases it may be felt that the argument remains sufficiently robust even in the presence of the challenge that has occurred. The process for assessing the impact of changes on the safety case is outside of the scope of this document, however the rationale for any such judgment should be explicitly documented, and itself be subject to review. Where updates to the safety case are required, our method can be applied in order to refine those aspects of the argument (returning to Stage 1).

Measures may be required to ensure the safety of the system is not compromised whilst the impact of the challenge is assessed and the necessary updates (to the system and the safety case as required) are put in place. This may for example require temporary restrictions on the operation of the system to mitigate the risk, or even a halting of system operations.

\section{Conclusions}
\label{sec:conclusions}

This paper has described a method for creating compelling safety cases. The method seeks to help improve safety case practice in order to address the weaknesses identified in current practice. In particular the method aims to address the issues of confirmation bias, after-the-fact assurance and safety case as paperwork exercise, that have been identified as key problems. Rather than attempting to create new notations and tools to address these issues, we contend that it is improvements in the safety case process that will make the most significant improvement to safety case practice. The new method described in this paper builds upon established approaches and best practice to create an approach that will ensure safety cases are risk-focused, seek to identify ways in which the system may \textbf{\textit{not}} be safe (rather than just assuming it is), drive safe design and operation of the system (influencing the system itself rather than just documenting what's there), are used to support decisions made throughout the life of the system, including system operation and change, and encourage developers and operators to think about and understand why their system is safe (and when it isn't). As systems and technology becomes more complex, such as through increasing use of autonomy and artificial intelligence, increased inter-connectivity and more dynamic and rapidly changing systems, improving safety case practice in this manner becomes even more important to ensure safety cases remain fit for purpose.

We have used a simple example of an infusion pump system to illustrate how the method is applied in practice. Further work to validate the method will focus on application to more complex and complete examples, and assessment of applicability by independent safety case developers.

\bibliographystyle{unsrt}  
\bibliography{bibliography}  

\end{document}